# Power Relations in Game Theory

Author: Daniele De Luca, University of Turin
E-mail: d.deluca@unito.it
Postal address: Via XXVII Marzo, 10093, Collegno (TO), Italy

*Abstract*

The concept of power among players can be expressed as a combination of their utilities. A player who obeys another takes into account the utility of the dominant one. Technically it is a matter of superimposing some weighted sum or product function onto the individual utility function, where the weights can be represented through directed graphs that reflect a situation of power among the players. It is then possible to define some global indices of the system, such as the level of hierarchy, mutualism and freedom, and measure their effects on game equilibria.

Along with the intuitive cases, where favoring someone improves the favored one's and worsens the favoring one's situation, there are many other counterintuitive cases where two players lose as they favor each other, or someone wins by being obedient, or two dominant nodes cancel their power setting free their servant.

## 1. Introduction

The aim of this paper is to analyze power relations in games. Following Steven J. Brams, "power is not much studied in game theory. In non-cooperative game theory, which does not assume a contract is binding, players are usually assumed to have an equal ability to influence the outcome."[1]

Shapley, L. S., and Shubik, Martin (1954), followed by many others, e.g. Allingham, Michael G. (1975) and Wiese, Harald (2009), have proposed indices of power in the context of Cooperative Game Theory. The Shapley value and the Banzhaf index measure, in different ways, the contribution of each player to each possible coalition. They are good candidates for the task, and became a basic concept for evaluating power in voting games.

This paper has a different and new approach. Rather than trying to analyze the power relationships present in the game as it is, we will apply different power structures that change the players' outcomes, resulting in a new equilibrium. Rather than applying Cooperative Game Theory to power relations, we will try to apply power relations to Non-Cooperative Game Theory.

It is common in economic and social sciences to think that individuals make decisions in order to maximize some utility function. However, the utilities of two individuals could coincide, even if only partially. It seems quite natural to express this possibility in terms of "influence" or "power": A has power over B because B acts partly for the sake of A, taking into account A's utility. Therefore the utilities of A and B are partially similar.

There is abundant literature on the subject of so-called "social preferences" that claims that human behavior can be influenced by social values such as fairness, equity and so on. While no one denies that possibility, the very weight of social preferences in human behavior is a controversial topic. See for instance Fehr, Ernst and Charness, Gary (2023) for an overall review and Binmore, Ken (2010) and List, John A. (2009) for more critical accounts of the field. As it will be clear in what follows, this paper does not consider universal preferences such as fairness, but only individual-specific preferences such as someone's willingness to favor a particular player. Furthermore, in the present context we will treat these preferences as mere data, voluntarily avoiding asking about their origins.

For some individual $i$ in a non-cooperative game with $n$ players, let us introduce the distinction between the inertial utility $u_i$ and the compound utility $U_i$. The former is the preference function that $i$ would have without any external influence; we take it as a datum.

[1] Dowding, Keith, ed. (2011): 270.

The latter is the result of the combination of inertial utilities of several players, in a certain order of preference.
Compound utilities can have infinite different forms. Two of the simplest are:

[1] $$U_i = \sum_{k=1}^{n} c_{k,i} u_k ,$$

[2] $$U_i = \prod_{k=1}^{n} u_k^{c_{k,i}}.$$

While in the additive formula [1] the inertial utilities of the players are perceived by the individual $i$ as substitutable, in the multiplicative formula [2] they are complementary, i.e. they take on greater value if they are pursued in a certain combination. Both [1] and [2] present a series of coefficients $c_{i,j}$ and it is convenient to think of them as if they have a constant sum: $\sum_{k=1}^{n} c_{k,i} = 1$ and $c_{i,j} > 0$. These $c_{i,j}$ weights determine the order of preference among the inertial utilities in some $U_i$. They are obviously independent from the chosen function, because they form the basis of any type of compound utility. To fix these coefficients it is worth defining the concept of "power system".

## 2. Power systems

A *power system* $\Gamma$ is completely determined, up to isomorphisms, by a set of nodes $V$ and a function $\pi$ which associates to an ordered pair of nodes the weight of their power relation (or directed edge).

$$\Gamma = (V, \pi)$$

$V$ is the set of the $n$ nodes of the graph,

$$V = \{v_1, v_2, \dots, v_n\}$$

while $\pi$ is a function defined as follows:

$$\pi: V \times V \rightarrow R,$$

$$(v_i, v_j) \longmapsto f_{i,j} \qquad (i \neq j)$$

[3] $$0 \leq f_{i,j}$$

[4] $$\sum_{k=1}^{n} f_{k,i} < 1.$$

Given any two nodes in the system $i$ and $j$, $f_{i,j}$ is 0 if there is not any power relation from $i$ to $j$, otherwise it has a value between 0 and 1 (both excluded) if such a relationship does exist. An equivalent way to represent the function is to use the so-called *weighted adjacency matrix*, i.e. the matrix $F \in R^{n,n}$ with outries $f_{i,j}$.

Now, we want to know how much power a node has over another once we have taken into account the whole structure of the directed edges.

Suppose that $i$ influences $j$ and $j$ influences $k$. This means that somehow $i$ must influence $k$, because it is likely that the relation of influence acts in a transitive way. For instance, if $i$ controls half of $j$'s behavior ($f_{i,j} = \frac{1}{2}$), and $j$ controls half of $k$'s behavior ($f_{j,k} = \frac{1}{2}$), how much will $i$'s influence be on $k$? It is quite evident that if half of $k$'s behavior is influenced by $j$

, half of this half is also influenced by $i$ (so $\frac{1}{2} \cdot \frac{1}{2} = \frac{1}{4}$). The transitivity of weights seems to imply a recursive multiplication.

French Jr, John RP (1956) calls this property *indirect influence*, when power “is exerted on another through the medium of one or more other persons.”[2] Among the theories of social influence, there is a specific tradition of models that Flache, Andreas, et al. (2017) call *models of assimilative social influence* that take into account transitivity of influence networks through time. For example, multiplying by itself the $F$ matrix we obtain subsequent steps in the flow of power relations.

We follow a different line. We want to determine the transitive property in a formal way without specifying a time variable. We therefore introduce a new matrix, the *colonization matrix* $C$:

$$C \in R^{n,n}.$$

The value $c_{i,j}$ in the $i$-th row and in the $j$-th column indicates the "colonization" of $i$ in $j$, and corresponds to the amount of influence that ultimately (once all the edges have been covered) $i$ has on $j$. If $c_{i,j} > 0$ then we say that $i$ *colonizes* $j$. In other words, $c_{i,j}$ indicates how much the inertial utility of $i$ affects the compound utility of $j$, defining its weight in the $U_j$ formula (see formulas [1] and [2]). As in the previous example, even if there is no edge from $i$ to $k$ ($f_{i,k} = 0$), $i$ colonizes $k$ ($c_{i,k} = \frac{1}{4}$).

The set $s_i = \{c_{1,i}, c_{2,i}, \cdots, c_{n,i}\}$ of colonizations of all nodes (including $i$ itself) in node $i$ therefore contains the coefficients of the compound utility function $U_i$.This set is nothing more than a column of the matrix $C$ which we will call the $i$’s "spectrum" and which represents how much $i$ is influenced by other nodes. Let us then introduce the following axioms:

[5] $$\sum_{k=1}^{n} c_{k,i} = 1$$

[6] $$c_{i,j} = \sum_{k=1}^{n} f_{k,j} c_{i,k} \qquad (i, k \neq j).$$

[2] French Jr, John RP (1956): 183.

[6] allows us to calculate[3], together with [5], the matrix $C$ from the matrix $F$. It is a recursive formula, because each colonization requires the calculation of the predecessor nodes recursively. The recursion stops when it reaches the colonization of a node in itself. [6] basically means that the spectrum of a node is occupied by the spectra of its predecessors in a proportion determined by the weight of the edges pointing to the node.

Axioms [6] and [5] allow to derive $C$ starting from any adjacency matrix, through a linear system of $n^2$ equations in $n^2$ variables. The function $\varphi$, that associates a colonization matrix with an adjacency matrix, is neither injective nor surjective, but it is sufficient to restrict the domain and the codomain of $\varphi$ to make the transformation reversible. If one excludes self-loops from the domain, i.e. edges that enter and exit the same node and form the diagonal of the adjacency matrix, and the codomain is restricted to the range, then a given colonization matrix allows for the construction of a non-homogeneous system in $n \cdot (n - 1)$ variables with which the respective adjacency matrix can be obtained. There is, consequently, a bijective function $\varphi_1$ from the set of adjacency matrices without the diagonal to the set of restricted colonization matrices, which associates each $F - d$ to one and different $C$, according to [5] and [6].

The representation of a 5-node power system is illustrated in *Figure 1*.

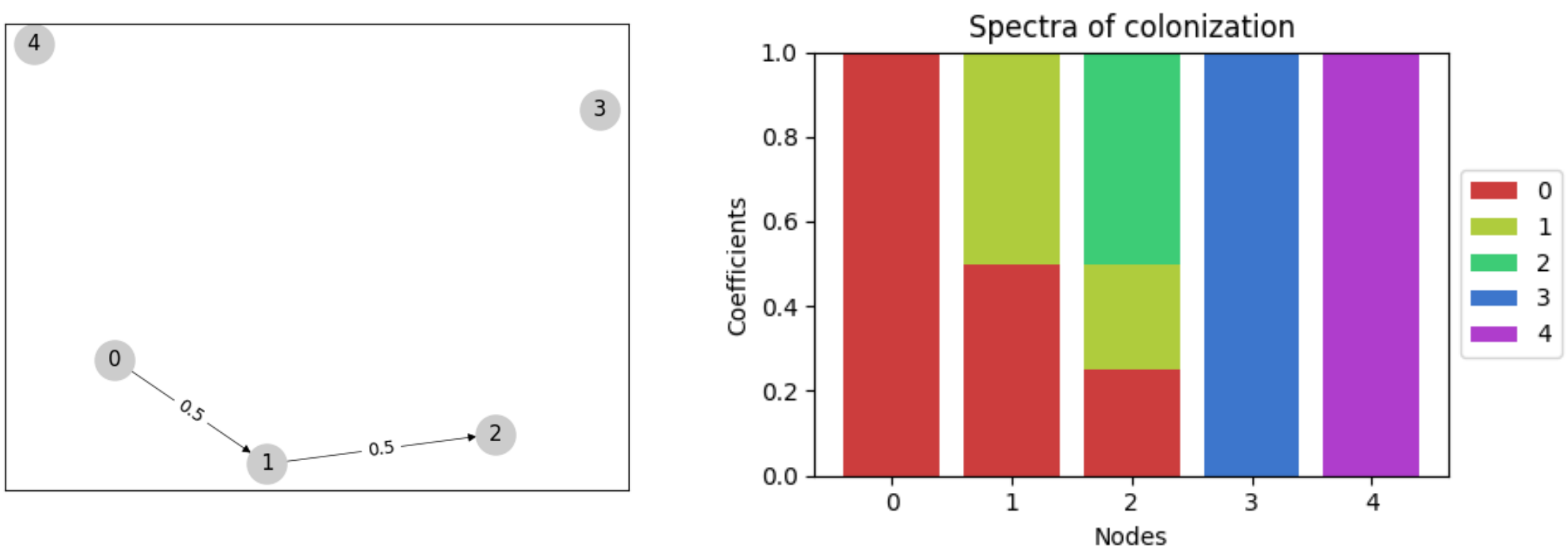

**Figure 1. a)** *graph that represents the power relations* **b)** *histogram that represents the colonization relations.*

It shows that (0) colonizes (2) not directly, but through (1): in *Figure 1b* the spectrum of (2) is a column of colors occupied for a quarter by the red color, which represents (0), and can be checked in the legend on the right. For this reason colonization is transitive, albeit in a gradually decreasing way.

Two types of relations have therefore been distinguished: 1) the *power relation*, represented by the graph and the adjacency matrix, which is neither transitive nor reflexive; 2) the

---

[3] The authors have created an open-source Python library, which can be found at https://pypi.org/project/power-sysGraph/, that calculates the colonization matrix corresponding to any adjacency matrix. All the figures in this paper were generated using that library.

*colonization relation* which, on the contrary, is reflexive (because each node has at least a part of the spectrum of its color, for {2}) and is transitive.

The first relation is represented by the matrix $F$ and the graph of *Figure 1a*, the second by $C$ and the histogram of *Figure 1b*. Since the columns of $C$ have a constant value (one), this matrix can, in fact, be represented in a histogram with constant height segmented bars.

A relationship that is reflexive and transitive is called a *preorder*. While it is called an *order* if, in addition to the aforementioned properties, it also possesses the antisymmetric one, which says that if $i$ is colonized by $j$ and $j$ by $i$, then $i$ is $j$. The antisymmetric property basically prohibits the formation of circularity. Colonization is therefore a preorder[4], but it is not an order, because it is not antisymmetric: a node can colonize a different node and be colonized in turn, so circularity is allowed. As we will see, however, in hierarchical systems the colonization relationship is an order because this type of system is always acyclic.

[4] See *Theorem 1* and *Theorem 2* in Appendix.

## 3. *Basic properties of nodes*

The colonization matrix allows for immediately fixing some basic properties of a node. Let us define the sum of row $i$ of matrix $C$ as the *total power* of the node $i$, and we will call this value $p_i$.

$$\sum_{k=1}^{n} c_{i,k} = p_i$$

In the histogram of *Figure 1b,* the total power of a node represents the sum of all the parts of the colonization spectra occupied by its color; it measures how much a node is effectively capable of influencing itself and others. For greater clarity, it is useful to represent the nodes in the graph with diameters proportional to their total power $p_i$.

On the diagonal of matrix $C$, where $i$-th row and $i$-th column intersect, we find a value that can be defined as *freedom* of the node $i$, which is the colonization of the node in itself. We can think of it as the probability that $i$ makes its own inertial decision, represented in the spectrum by the portion of the same color of the node in the node bar.

$$l_i = c_{i,i}$$

A node is completely free if, and only if, it has no incoming edges[5].

[5] See *Lemma 1* in Appendix.

## 4. *Two nodes relations*

In this section we introduce some indices in order to analyze the relation between two nodes only. Later, we will extend these indices to relations among $n$ nodes.
Cycles are allowed, so two nodes can have power over each other. In this case the nodes collaborate, or rather, they *co-influence* or *mutualize*, thus establishing reciprocity. The following definition gives a method to measure this property.

Definition (1): *mutualism* between $i$ and $j$ is twice the minimum between the colonization of $i$ in $j$ and that of $j$ in $i$.

$$mutualism(i,j) = 2 \cdot min(c_{i,j}, c_{j,i})$$

The reason for the "minimum" is that we want to select only that part for which the colonizations are reciprocal, therefore the common part of two lengths. The reason for the "double" is that we are counting that part, the lesser value, on two spectra.
The term "mutualism" was chosen instead of "cooperation" because the latter is intrinsically ambiguous. It is true that we often speak of "cooperation" when the responsibility for decisions is shared by all the members of a group, therefore we use it as a synonym for "reciprocity". However, "cooperation" literally means that individuals (or groups) act together for a common purpose. The purpose could also have been decided by only one of the individuals, it could consist of the advantage of only one of them. In this sense, even dictatorships or slavery rely on cooperation. Indeed they are almost perfect forms of cooperation: the master and the slave work together, in perfect agreement, for the sake of the master alone.
In more formal terms, *cooperation* will have to be defined in a different way from mutualism, i.e. by measuring the similarity between the purposes of the two nodes.

Definition (2): *cooperation* between $i$ and $j$ is the sum of the colonizations of $i$ in $j$ and of $j$ in $i$.

$$cooperation(i,j) = c_{i,j} + c_{j,i}$$

Cooperation increases by adding any edge; in this sense it is more a measure of the strength of the connections. Mutualism, on the other hand, depends only and solely on reciprocity: it can also be used as a general measure of circularity strength within directed graphs. There can be no mutualism without cooperation, but the reverse is not true.
Let us define another measure of the relationship between two nodes: hierarchy.

> Definition (3): *hierarchy* between $i$ and $j$ is the difference (in absolute value) between the colonization of $i$ in $j$ and that of $j$ in $i$.

$$hierarchy(i, j) = \left| c_{i,j} - c_{j,i} \right|$$

Hierarchy measures the disparity in power between two nodes[6]. In the absence of edges it is always zero, as well as if the two nodes colonize each other in equal measure.
As we said before, there is no hierarchy without cooperation, even though there can be cooperation without hierarchy.

---

[6] "It is the fundamental idea and key characteristic of hierarchy that privileges and prerogatives are allocated unequally amongst members of the social system according to a system of social rank—whatever the specific criteria this is based on." Diefenbach, Thomas (2013): 37.

## 5. N-nodes relations

Mutualism, cooperation and hierarchy are relations. This means that they are not concerned with a single node (such as total power and freedom), but with a couple. They are also symmetric. The order in which the pair is inserted does not matter: the measure of hierarchy $(i, j)$ will always be the same as that of hierarchy $(j, i)$. The simplest way to measure these properties for groups of more than two nodes, thus creating indices, is to extract from $V$ (the set of nodes) all the subsets of cardinality 2, i.e. the pairs (not ordered) of different elements. Let us call this set of combinations $V_2 = \{X \in P(V) : |X| = 2\}$, where $P(V)$ is the power set of $V$. As long as $V$ has cardinality $n$, the cardinality of $V_2$ is $\frac{n!}{2!\,(n-2)!} = \frac{n^2-n}{2}$. The maximum value that the three sum-functions can take on of $n$ nodes will be $n - 1$, so we can normalize dividing them by that value.

$$mutualism(\Gamma) = \frac{\sum_{\{i,j\} \in V_2} mutualism(i,j)}{n-1}$$

$$cooperation(\Gamma) = \frac{\sum_{\{i,j\} \in V_2} cooperation(i,j)}{n-1}$$

$$hierarchy(\Gamma) = \frac{\sum_{\{i,j\} \in V_2} hierarchy(i,j)}{n-1}$$

Mutualism measures the circularity of the system, because only through cycles can colonizations become reciprocal. Hierarchy is an opposite measure, that is, of the amount of acyclic connections, while cooperation simply measures the connectivity of the system, be it cyclic or acyclic.
The relation among these three indices is expressed by the following formula[7]:

$$hierarchy = cooperation - mutualism.$$

It is therefore sufficient to know two of these properties of a system to derive the third.
Finally, one could define the total freedom of a system as the average of individual freedoms. However, this average would not go from 0 to 1 (excluded) like the other indices, but from $\frac{1}{n}$ (excluded) to 1, because in every system there must be more than the equivalent of one free node[8]. Let us therefore normalize the sum of individual freedoms in a new index from 0 (excluded) to 1 and call it the total *freedom*[9] of the system.

[7] See *Theorem 3* and *Theorem 4* in Appendix.

$$freedom(\Gamma) = \frac{\sum_{i=1}^{n} c_{i,i} - 1}{n-1}$$

In every system:

$$freedom = 1 - cooperation.$$

Finally, by substituting in the previous theorem, we obtain that:

$$hierarchy + freedom + mutualism = 1.$$

These theorems apply to any transitive influence network and therefore have general validity. In fact, any system of this type must have weights that indicate who influences whom and to what extent, and therefore must contain a colonization matrix, which is the only requirement for calculating the indices.

Thus we introduce a typology of systems: a system is 1) *hierarchical* when it is devoid of mutualism and its hierarchy is greater than zero (*Figure 2c*), 2) *mutual* when its hierarchy is zero and its mutualism is greater than zero (*Figure 2b*), 3) *free* when its cooperation is zero and it is therefore made up of isolated nodes (*Figure 2a*).

In a hierarchical system the colonization relation is always a *(partial) order* because it is antisymmetric, given the previous definition.

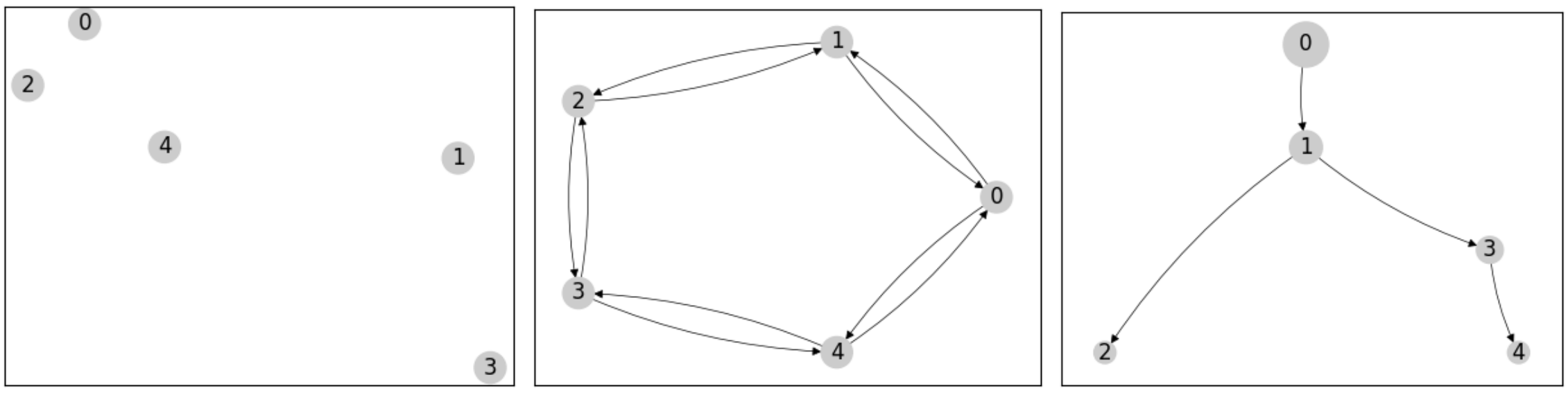

**Figure 2.** *Examples of 5-node systems: **a)** free system **b)** mutual system **c)** hierarchical system.*

[8] This is a simple consequence of *Theorem 5*.
[9] See *Definition 7* in Appendix.

Since the sum of freedom, mutualism and hierarchy is always one, these properties can be represented with a pie chart or as in this case, a donut chart. In *Figure 3* you can see some power systems along with their properties.

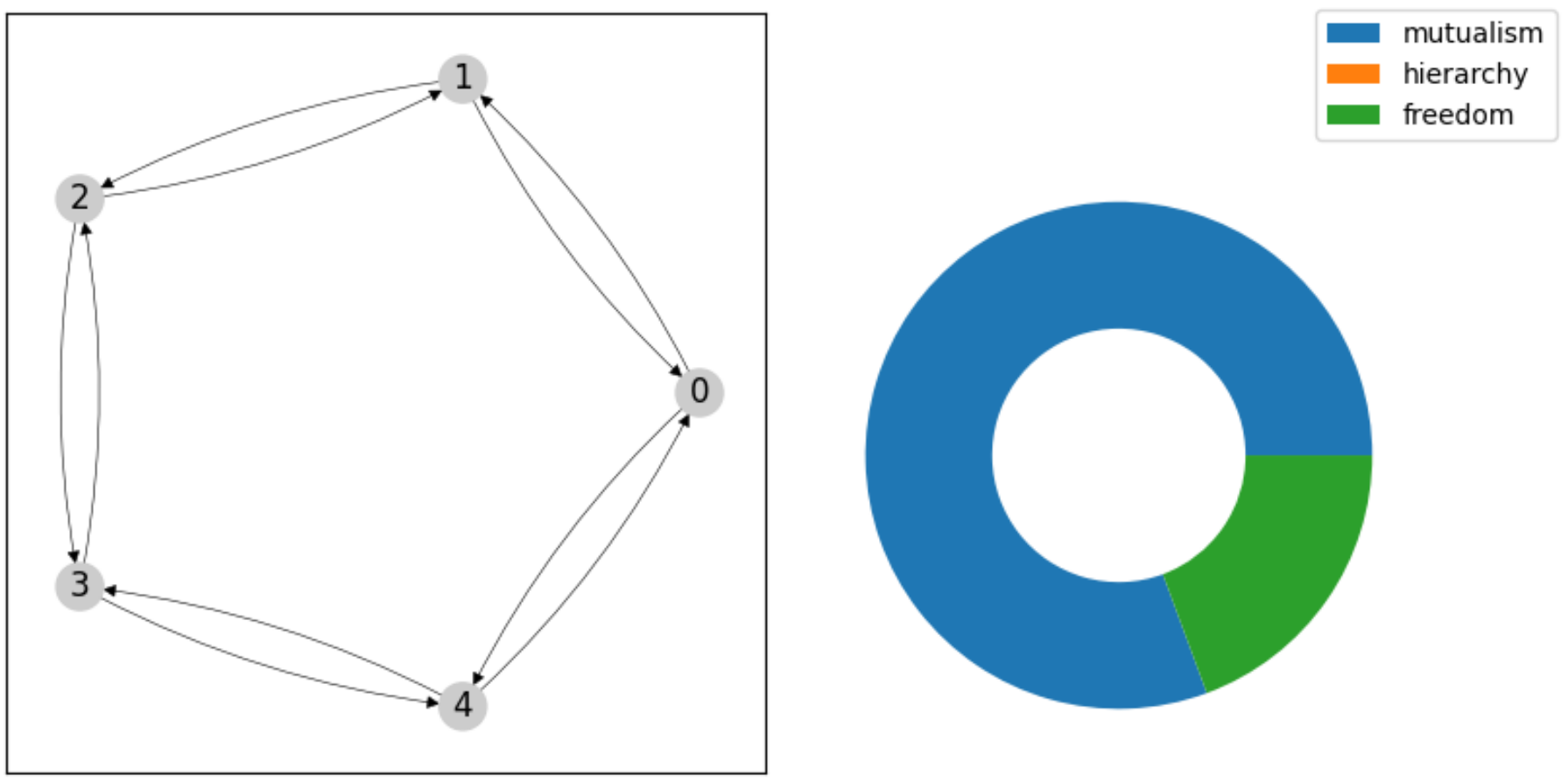

**Figure 3.1**

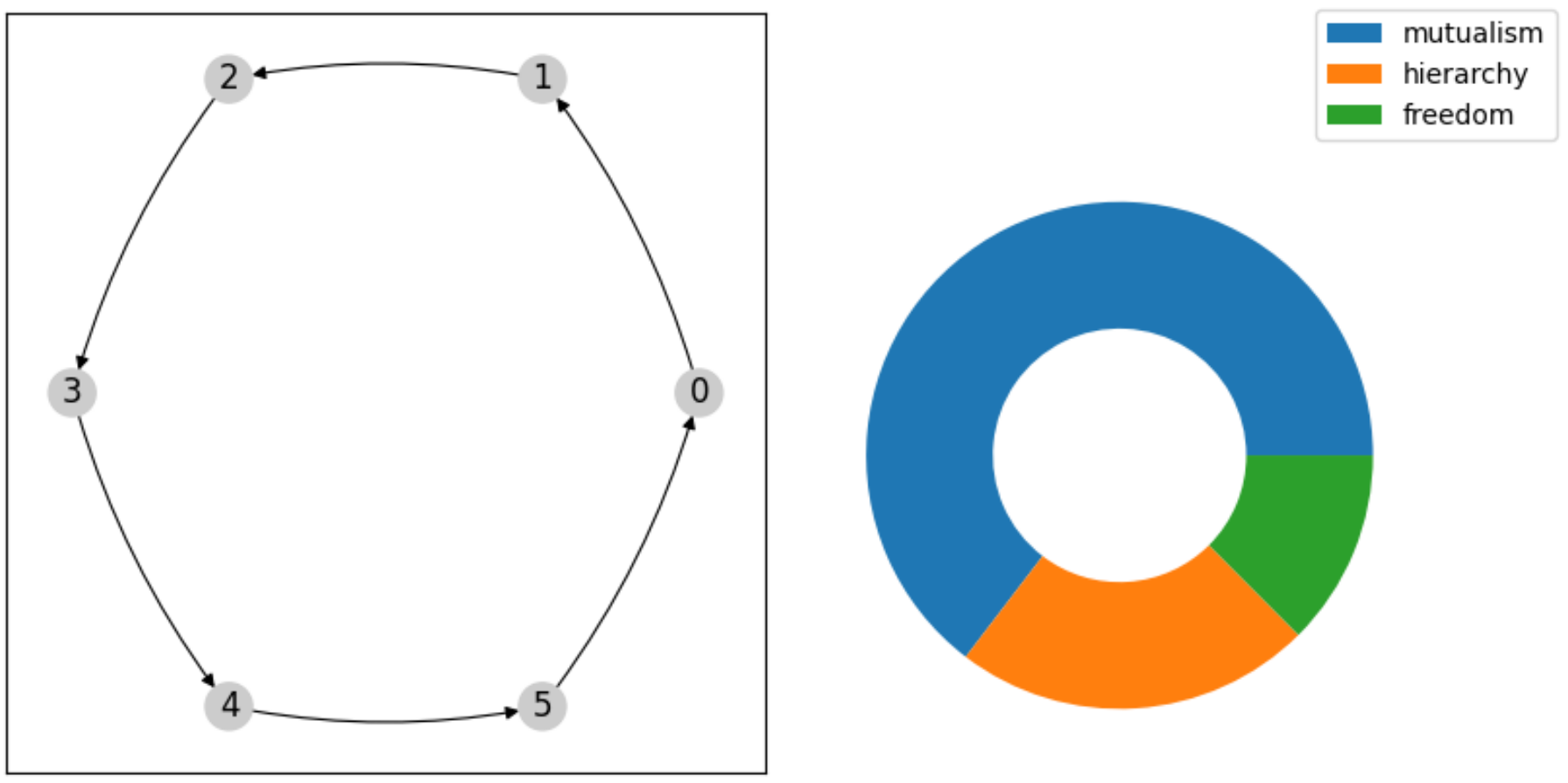

**Figure 3.2.** *Hierarchy in a one-way cycle is proportional to its length, while mutualism is inversely proportional*[10].

[10] This interesting property of our colonization-based hierarchy measure makes it different from most of those that have been proposed in literature. See for example Mones, Enys (2012), Czégel, Dániel and Palla, Gergely (2015), or Moutsinas, Giannis (2021). Because most hierarchy measures are currently based on level distinctions among nodes, they cannot detect hierarchy in one-way cyclic situations, where all nodes lay at the same level. Yet, it is clear that in a large one-way cycle the relation between neighboring nodes *i* and *j* cannot be mutual, because the two paths from *i* to *j* and from *j* to *i* have different lengths. From the perspective of a single node, a very large one-way cycle is almost indistinguishable from a chain.

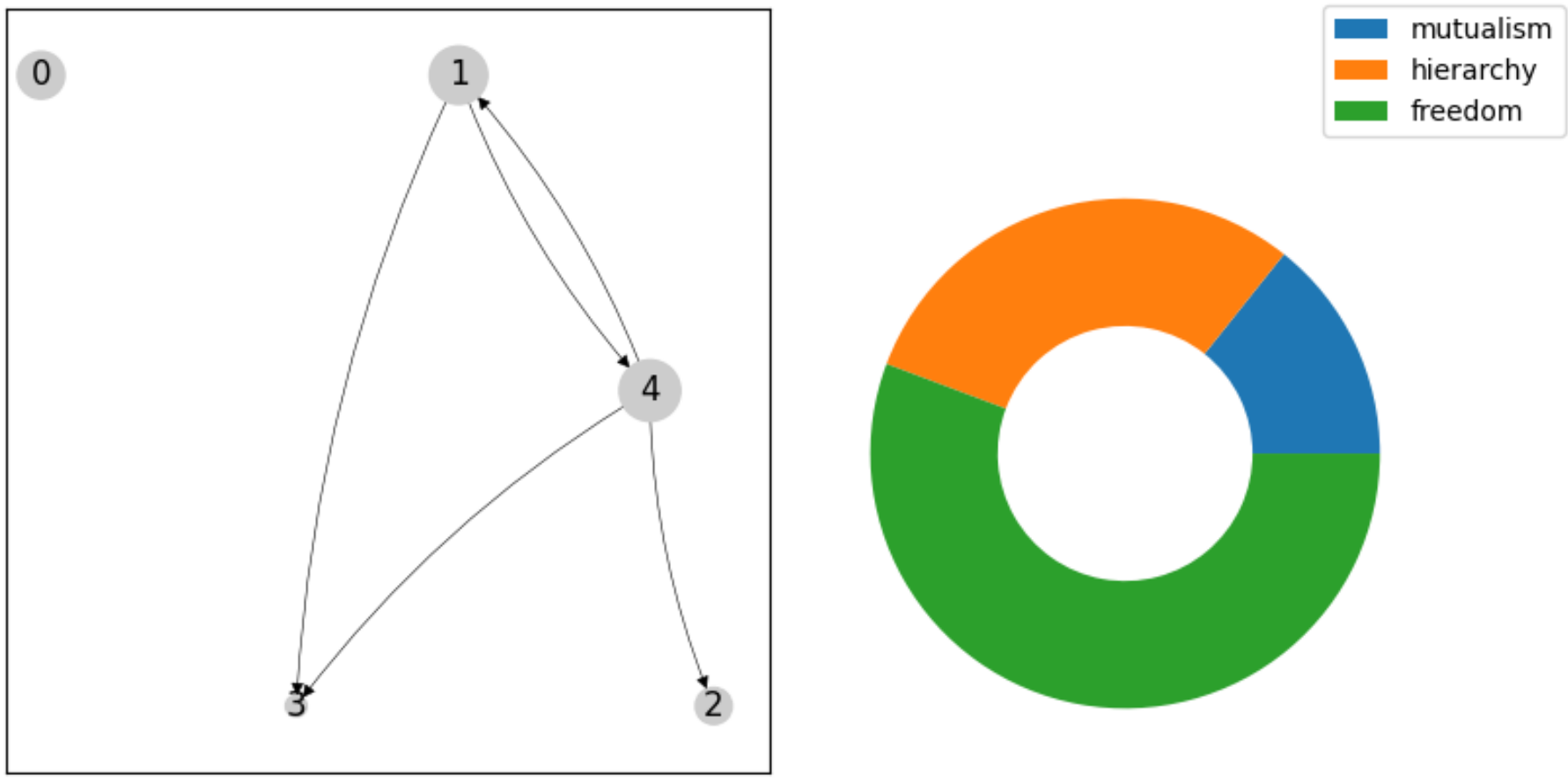

**Figure 3.3**

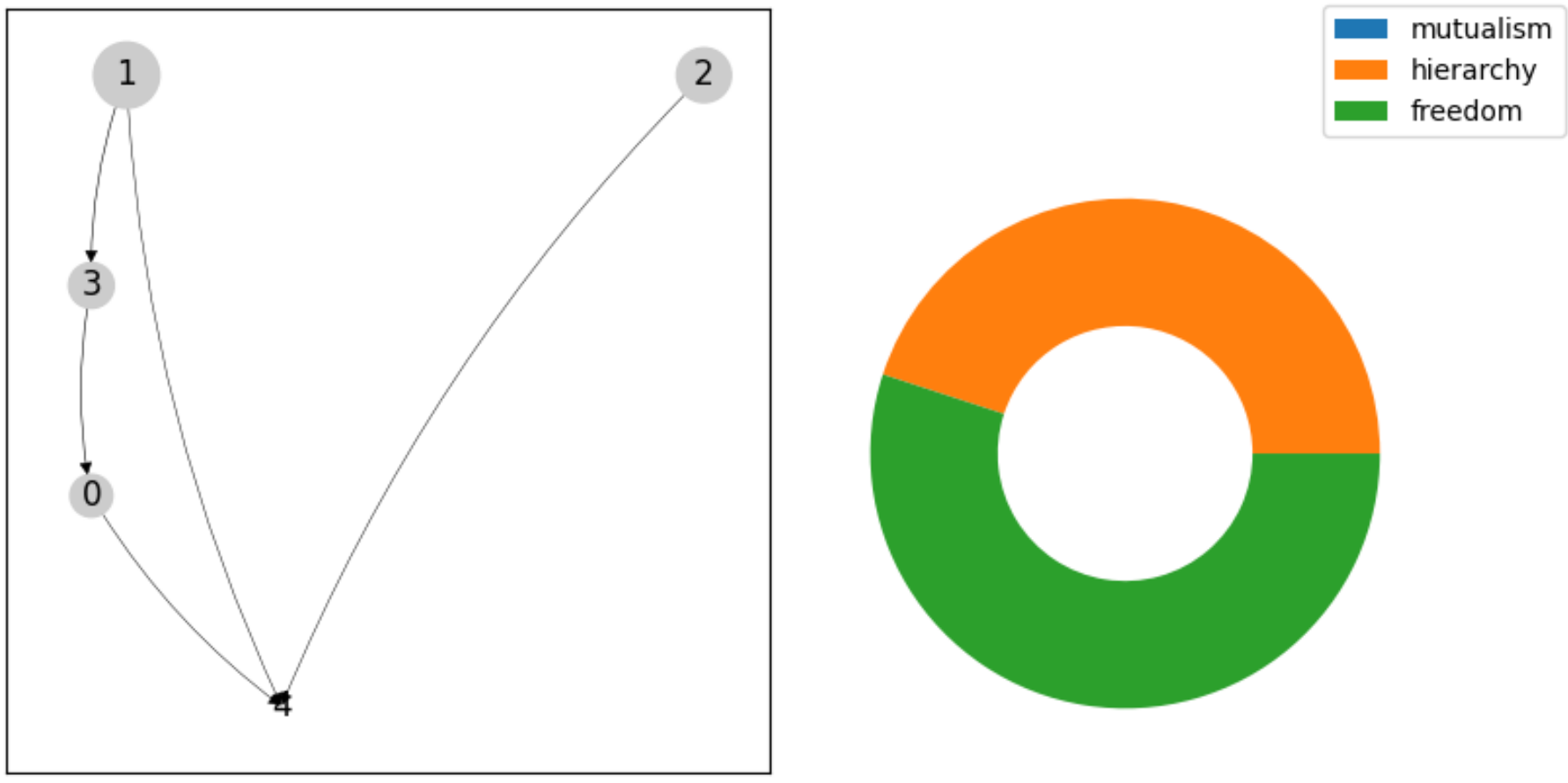

**Figure 3.4**

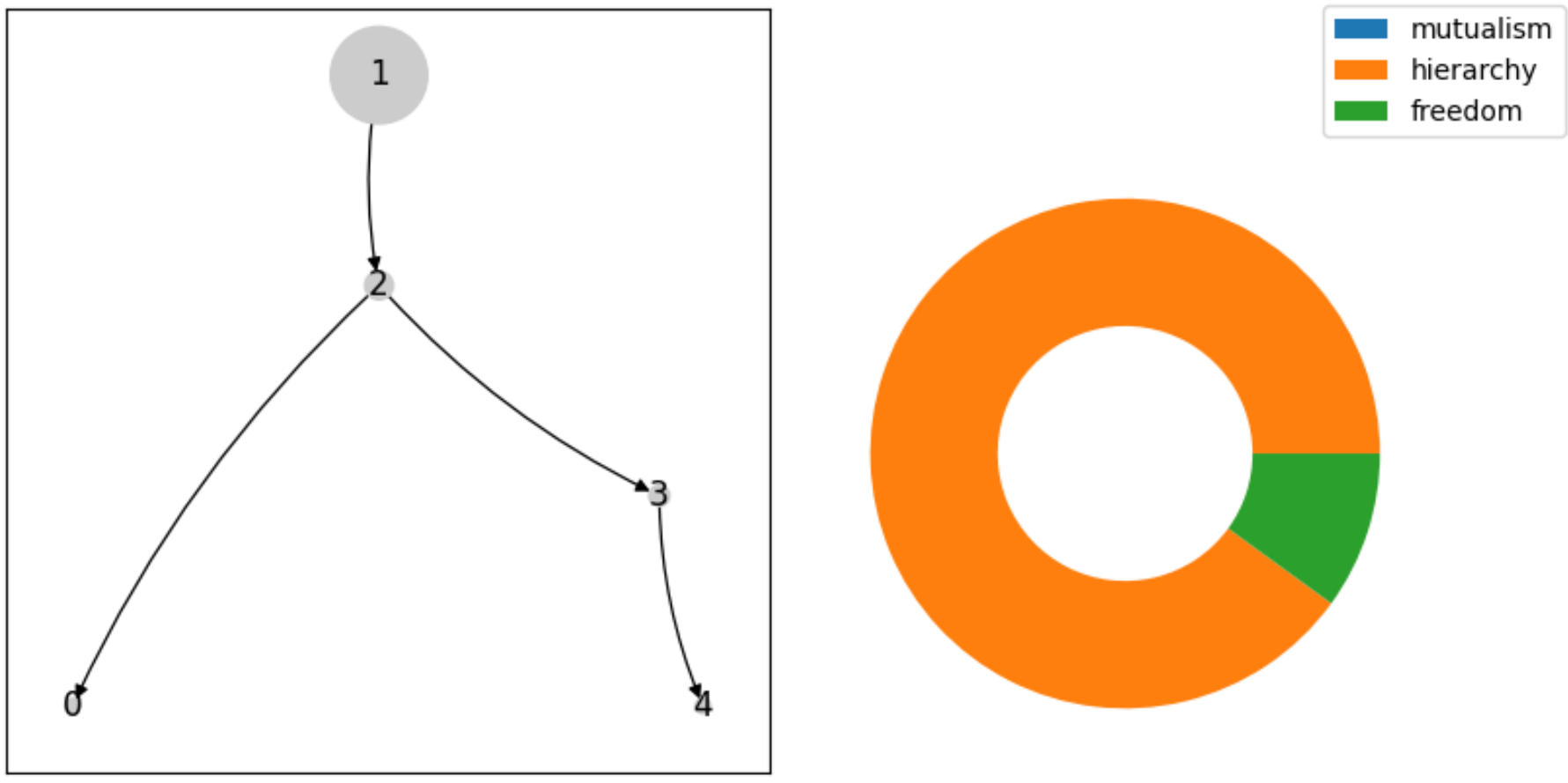

**Figure 3.5**

*6. The Prisoner's Dilemma*

The interest in the Prisoner's Dilemma lies largely in the paradox that, if we believe that players are rational decision-makers in the traditional sense, that is, that each of them always acts to maximize an individual utility, then they will both choose Defects. This is despite the fact that evidently the choice Cooperates would be more advantageous for all.

| PRISONER'S DILEMMA (*Table 1*) | | Player 2 | |
|---|---|---|---|
| | | Cooperates | Defects |
| Player 1 | Cooperates | -1, -1 | -6, 0 |
| | Defects | 0, -6 | -5, -5 |

The table above illustrates this situation. In each single box, the scores on the left relate to player 1, those on the right to player 2. For both players, the Cooperates choice is strictly dominated, meaning it is always worse compared to Defects, regardless of the choice made by the opponent.

The game is well known. Now we will try to apply different systems of power to the game. Players will obviously be represented by two nodes. In the absence of edges, each node follows the inertial decision and therefore the Nash equilibrium is Defects-Defects.

Yet what happens if one applies different power systems? In this case the payoffs of nodes will have to take into account the changes undergone by their colonization spectrum. Suppose for example that the colonization of (2) in (1) is worth 0.5. According to the definition we have given for colonization, (1) will change the compound utility function, which now also takes into account (2). These scores will be half colonized by player (2)'s scores. In the following tables we have used [1], i.e. the additive compound utility formula, but with formula [2] the results differ slightly, as long as we avoid negative and close to zero payoffs[11].

We can see the transformation of (1)'s payoffs in the following table.

| PRISONER'S DILEMMA | Player 2 |
|---|---|

[11] Remember that what matters in the payoffs is not their absolute, but their relative value.

| (*Table 2*) | | Cooperates | Defects |
|---|---|---|---|
| Player 1 | Cooperates | -1, -1 | -3, 0 |
| | Defects | -3, -6 | -5, -5 |

The new Nash equilibrium that is being determined has been marked in red. As we would expect, the player who undergoes colonization acts now in favor of the other. The conditions have not changed, of course: here (1) does not risk a different number of years. The power we deal with cannot change reality; it concerns only and solely the players' choices. The fact is that our choices do not depend just on material conditions, as they involve a subjective perception of those conditions.

On the other hand, this lies at the foundation of the theory of marginal utility: it is not the value in itself (assuming it exists) of the commodity that makes me decide at what price I am willing to pay for it, but the value *for me*, specifically, what value I perceive in it. Utility cannot be measured objectively, because it is an intrinsically subjective property. This also applies to the classical individual function: the pattern of *Table 1* is also subjective.

In other words, for (1) those 6 years that the player will have to pay (see Table 2 and the Cooperates-Defects box) are worth 3, (1) perceives them as less burdensome. This is perhaps because one gains in having a clear conscience, in being altruistic, or simply in avoiding punishment. The reason does not matter, nor does it matter how many years one will actually have to pay for the purposes of his/her choice. What matters is only his/her perception of the utility of that choice.

It is often said that the utilities of different individuals are not comparable, or that the utility function is not of interest for its cardinal value, which could be related to that of another individual, but only for an ordinal one. To merge various utility functions, however, it is not necessary to think of their supposed cardinal value because, even the spectrum, as well as inertial utility, can be interpreted in purely subjective terms. It is *for me* that your utility has that value in relation to mine, and it does not matter if this is objectively true.

As a third possible situation, let us now take the case in which two nodes mutually condition each other with edges of range 0.5 as in *Figure 4*.

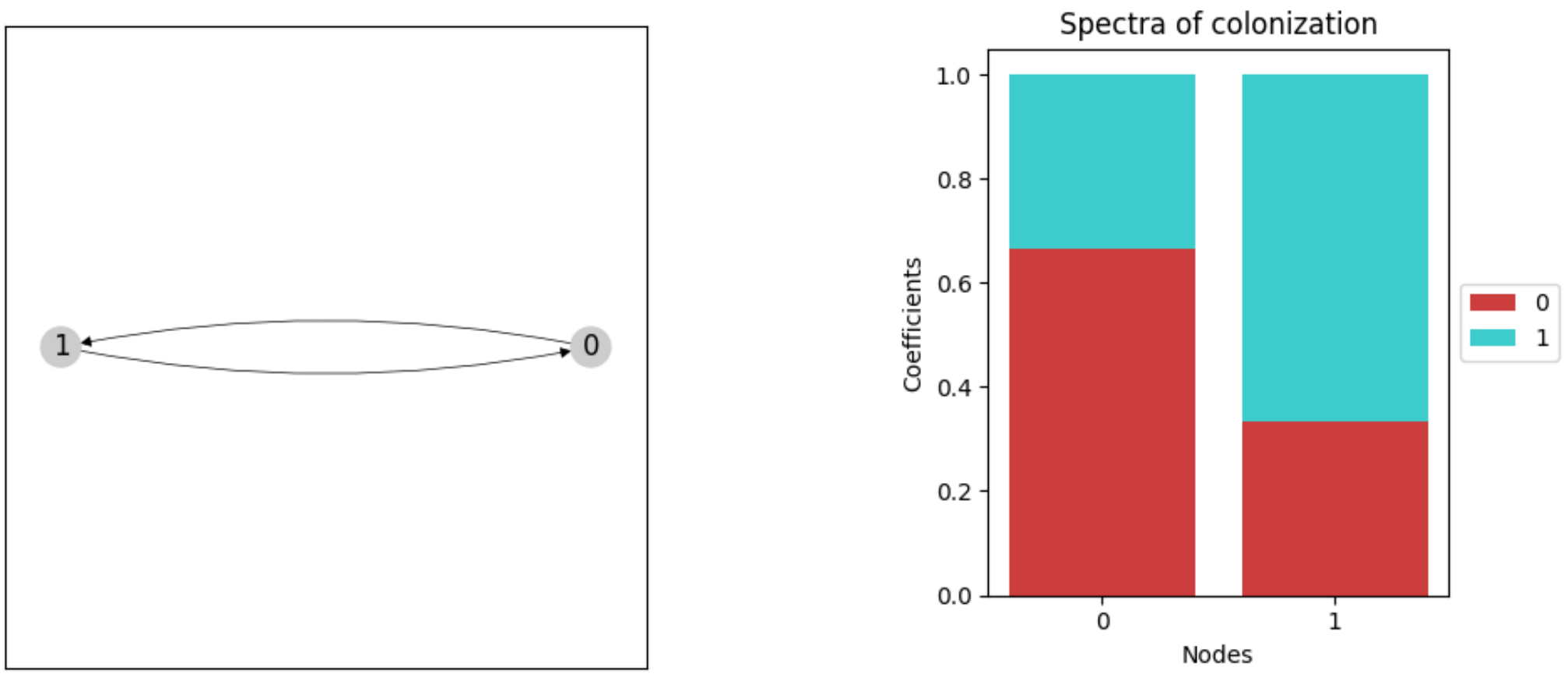

**Figure 4.** *2-node system with two mutual 0.5-weighted edges:* ***a)*** *adjacencies* ***b)*** *colonizations.*

Here the colonization of each node in the other is $\frac{1}{3}$. Now the scores of both players have changed and the Nash equilibrium turns out to be Cooperates-Cooperates.

| PRISONER'S DILEMMA (*Table 3*) | | Player 2 | |
|---|---|---|---|
| | | Cooperates | Defects |
| Player 1 | Cooperates | -1, -1 | -4, -2 |
| | Defects | -2, -4 | -5, -5 |

Unsurprisingly, mutualism allowed the two prisoners to reach a more favorable score for both.

In fact, there is no doubt that *Table 3* corresponds to a possible outcome of the story described. Rarely would two people who know each other, having committed or not a crime together, act without taking into account the payoffs of the other. It should be noted that in the traditional solution to the prisoner's dilemma, our choice is completely independent of the other's scores. The Nash equilibrium depends only and solely on our score[12]. However, in the context of human relationships, this type of complete absence of interest in the other's utility or harm (and in the other's actions towards us) is quite improbable[13].

[12] This is true in the case of the Prisoner's Dilemma because the Cooperates strategies are strictly dominated, so they are eliminated regardless of the choice of the other player.

[13] See for example Peterson, Martin, ed. (2015): 243, where an experiment is presented on a game that requires a mutualism index of 28%. Even in the less favorable situation for mutualism, that is, in single rounds between strangers, the participants had an average of 29% collaborative choices. In more favorable situations this value reached up to 84%. "And yet, among the most robust results to come out of experimental economics is that human players do not behave as game theory predicts. In laboratory Prisoner's Dilemmas, individual contributions toward the group good can be substantial."

The question then becomes, given any Prisoner's Dilemma, what level of mutualism must the two players possess in order not to accuse each other?

## 7. *Generalized Prisoner's Dilemma*

The first step is to generalize the Prisoner's Dilemma. Let us substitute payoffs with variables.

| PRISONER'S DILEMMA (Generalized) | | Player 2 | |
|---|---|---|---|
| | | Cooperates | Defects |
| Player 1 | Cooperates | $p, p$ | $q, r$ |
| | Defects | $r, q$ | $s, s$ |

In general, a Prisoner's Dilemma situation occurs when $q < s < p < r$. We have seen in the previous tables that the power systems cannot change the scores of the diagonal of the table because, in the corresponding boxes, the players have identical payoffs. The difference is made by the gray cells, i.e. the variables $r$ and $q$. By "approaching" the payoffs of the two players in those boxes we can move to the situation where the dominated strategy for both players is Defects, as in *Table 3*.
Thus we arrive at the following theorem[14]:

> In the generalized prisoner's dilemma, the two players will both choose Cooperates, if and only if, $mutualism(\Gamma) > 2 \cdot max(\frac{r-p}{r-q}, \frac{s-q}{r-q})$.

Regarding the example set out in the previous chapter, a simple calculation shows us that the minimum mutualism required is 33%.
However, reciprocity is not always possible. Two individuals can only colonize each other for less than half of their spectrum; yet, if the mean of $r$ and $q$ is not between $s$ and $p$, reciprocity cannot be achieved by eliminating the strictly dominated strategies. Two distinct situations then arise. Recalling that $r > p > s > q$, if $p$ and $s$ are closer to $r$, then we have not one, but two Nash equilibria: Cooperates-Cooperates and Defects-Defects, the first having a more advantageous score than the second for both.
Let us take the following score table as an example.

| PRISONER'S DILEMMA | | Player 2 | |
|---|---|---|---|

[14] See *Theorem 5* in Appendix.

| (*Table 1.1*) | | Cooperates | Defects |
|---|---|---|---|
| Player 1 | Cooperates | -1, -1 | -6 (-3), 0 (-3) |
| | Defects | 0 (-3), -6 (-3) | -2, -2 |

In this table, the maximum level of mutualism is the one with the scores in parentheses. The Cooperates-Cooperates box now has higher scores than the neighboring one so it is an equilibrium.

On the other hand, the equilibria Cooperates-Defects and Defects-Cooperates are always achievable[15]. Hierarchy seems more efficient in changing the equilibrium even when the target is very harmful for one of the players, while with mutualism that would be impossible.

[15] See *Theorem 6* in Appendix.

## 8. *The Ecology Dilemma*

It is not difficult to imagine something like the Prisoner's Dilemma with more than two players. We will call the following, an instance of so-called *public goods games*, the Ecology Dilemma.

Let us imagine some land with $n$ inhabitants. Each inhabitant decides whether to plant a tree. There is only one kind of tree in this country which does not produce anything edible. Furthermore, for the inhabitant who plants it, the tree has a cost: the work of planting it and taking care of it and the space it occupies on the land. There are benefits to planting a tree, but they are shared equally by all the inhabitants of the country: cleaner air, less risk of landslides, increased diversity in ecosystems, and so on.

We will call $p$ the cost of a tree and $r$ the income it brings to each inhabitant. Naturally $r < p$ and for an inhabitant it is not profitable to plant a tree. Thus the Ecology Dilemma in non-cooperative Game Theory has a very simple solution: no one will ever plant a tree. But if we apply different systems of power to the game, the balance will change quite predictably.

In a free system each inhabitant's gain from planting a tree will be $r - p$, which is a negative number, while that of any other inhabitant will be $r$. In this case, individual colonizations are not necessary. We only have to distinguish the freedom of the inhabitant $i$ ($c_{i,i}$) from the sum of all the other colonizations in the spectrum: $c_\epsilon = \sum_{j \neq i} c_{j,i} = 1 - c_{i,i}$. Using the additive compound function [1], the perceived payoff will be:

$$U_i = c_{i,i}(r - p) + c_\epsilon r =$$
$$(1 - c_\epsilon)(r - p) + c_\epsilon r =$$
$$r - p + c_\epsilon p.$$

It is easy to see that when any inhabitant is colonized, from how many nodes it does not matter, beyond a certain limit value, which corresponds to

$$c_\epsilon > \frac{p-r}{p},$$

a tree is planted.

We start from a free system, where, as mentioned, no trees are planted. By increasing collaboration in the system, the minimum level is reached allowing for the first tree to be planted. Evidently, an inhabitant must have been colonized enough to decide to plant one. *Figure 5* shows a hierarchical system, and the level of hierarchy is equal to that of cooperation, just greater than $\frac{p-r}{p} \cdot \frac{1}{n-1}$. There are many graphs that have these characteristics. In all of them there is always only one inhabitant who is subdued by a single other inhabitant or by any number of other inhabitants. The result does not change because the nodes that do not participate act as *free riders*.

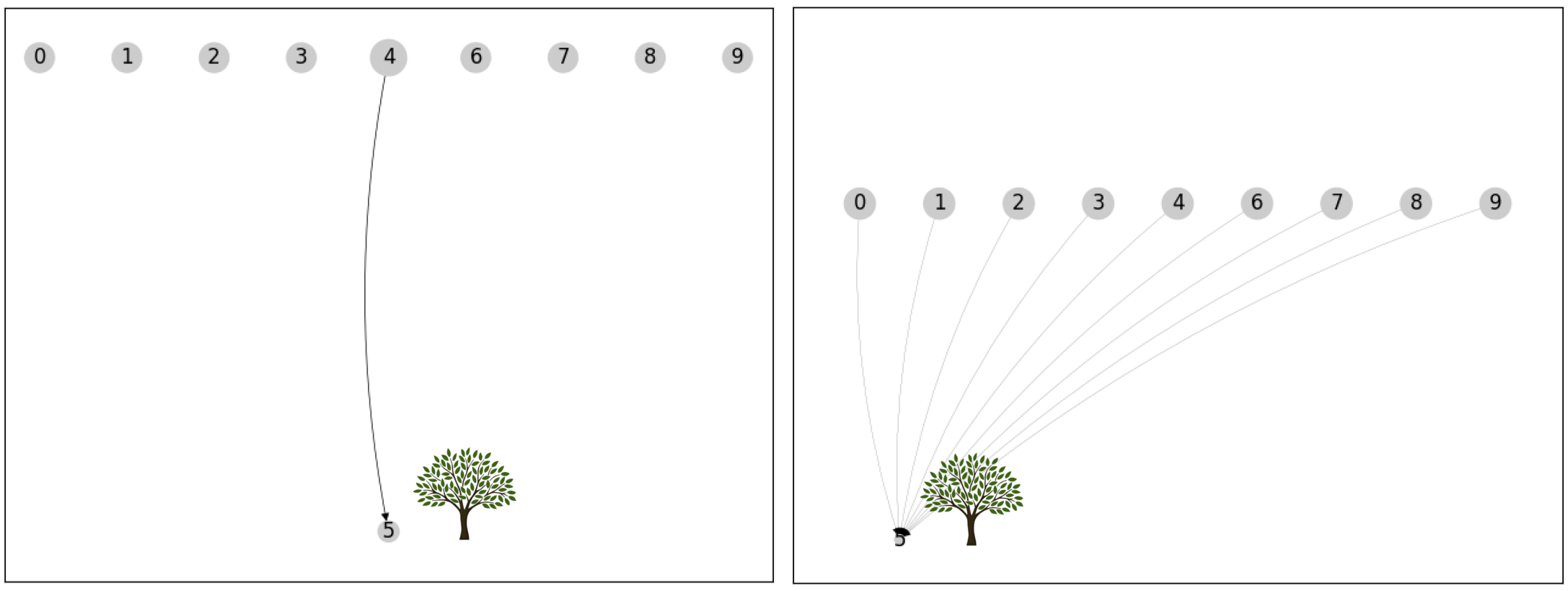

**Figure 5.** *The inhabitant (5) plants a tree:* ***a)*** *(5) is subjected to a single node,* ***b)*** *(5) is subjected to all the other nodes.*

Of course it does not seem like a great situation to be the only one to work for others. Yet our subdued residents do it, either because they are forced to, or because they identify with their neighbor, behaving in an "irrational" way. Again, what matters here is the result of power relationships, not their motivations.

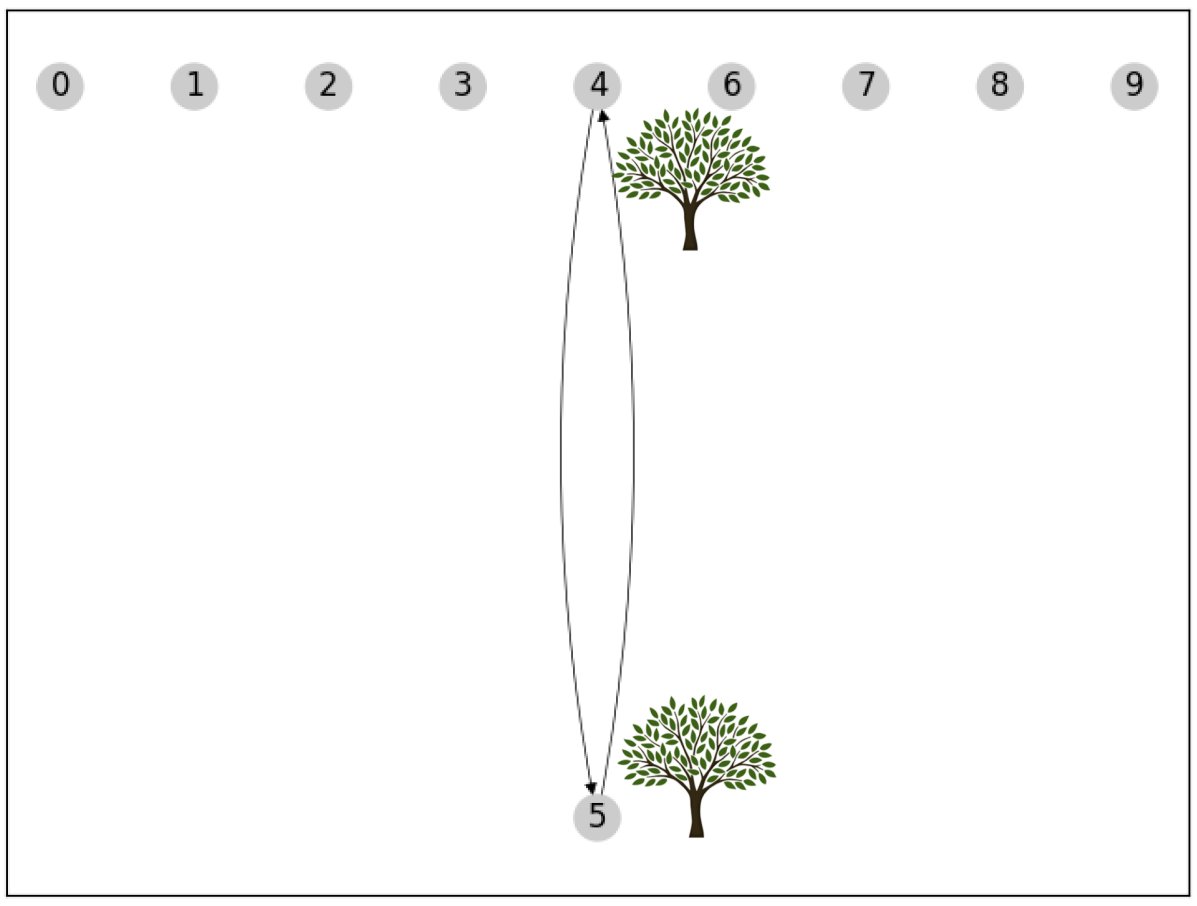

**Figure 6.** *Inhabitant (5) and ~~the~~ inhabitant (4) colonize each other.*

In *Figure 6* one can see the simplest mutual system. It is not possible for a single tree to be planted in a mutual system like this. The minimum is 2. The level of cooperation needed is exactly twice as before, $2 \cdot \frac{p-r}{p} \cdot \frac{1}{n-1}$, but one more tree is planted here.

In some cases there is not even the possibility of planting some trees by only two inhabitants in a mutual way, although a collaboration between larger groups of inhabitants is necessary. In particular if $2r < p$, that is, if the revenue is less than half the cost of a tree, two inhabitants are not enough to plant trees mutually. Nonetheless it is always possible, however high the cost and low the revenue, that an inhabitant decides to plant a tree in a hierarchical system. We have seen something similar in the previous chapter regarding the Prisoner’s Dilemma. An advantage of the hierarchy is to be able to push the submissive individuals into action even when the benefits for them are minimal.

But there is a downside. If we limit ourselves to a hierarchical system of two inhabitants, there is no way for them to plant *two* trees. Only *one* tree can be planted, by just one inhabitant or the other. In fact, in hierarchical systems, free riders are inevitable. This is even more surprising, if you think that hierarchy is often indicated as the only way to control free riders, because of its mentioned ability to push the submissive into action. Hierarchy can indeed control free riders on the bottom, yet at the price of allowing free riding on the top.

In a hypothetical Evolutionary Ecology Dilemma, imagine groups with different systems of power and generation after generation reproduce only the most suitable groups. It is not difficult to see scenarios in which the hierarchy is evolutionarily successful, but opposite scenarios as well, where mutualism supplants hierarchy.

9. *The Landowner Game*

Next game is a variation of the famous Cournot model.

The classic Cournot model concerns two or more firms competing. The firms decide how much product to bring on the market; if they compete, their profits decrease in favor of a third party, the consumer, who can buy a greater quantity of the commodity at a lower price. What happens in the system when we insert the consumer as a node?

Let us change the perspective. Now the agents who compete are peasants, who offer work in exchange for a wage, and the only consumer is a landowner. It is therefore a *monopsony*, because there is only one consumer and several producers (of work) competing. We will call it the Landowner Game.

Peasants decide how much to work for the landowner. Their hourly wage $W$ is set by this (inverse) demand curve:

$$W = a - Q$$

In this formula, $a$ (which in the examples below will be equal to 20) is the maximum amount of hours of work that the landowner would request if the peasants worked for free. While $Q = \sum_{i=1}^{n} q_i$ is the amount of work hours supplied by the peasants.

The inertial and compound utilities of workers are:

$$u_i = (W - C)q_i$$

$$U_i = \sum_{k=1}^{n} (c_{k,i} u_k)$$

where $C$ (cost) is the minimum wage a peasant could accept for work and will be 1 in the following examples. It should be noted that, as previously with the firms, it is possible that $u_i = 0$ e $q_i > 0$. In other words, the peasants could also work with zero profit because in this case $W = C$: the wage is the least that peasants could accept.

But what is the utility of the landowner? In fact, the landowner is interested in having more work for less money. Yet the demand curve relates labor to wages in such a way that the more labor, the lower the wage, so ultimately the landowner's utility can be set as simply proportional to the amount of work:

$$u_L = Q = a - W.$$

The consumer node, therefore, has a different utility function than the other nodes. Furthermore, in this simple model, the landowner cannot make any decisions.

A node that cannot make decisions is *passive*, but its inertial utility can affect *active* nodes. In this case only the peasants, the active nodes, can make a choice: how much work to offer to the landowner, the passive node. But their choice also depends on the power the landowner has over them.

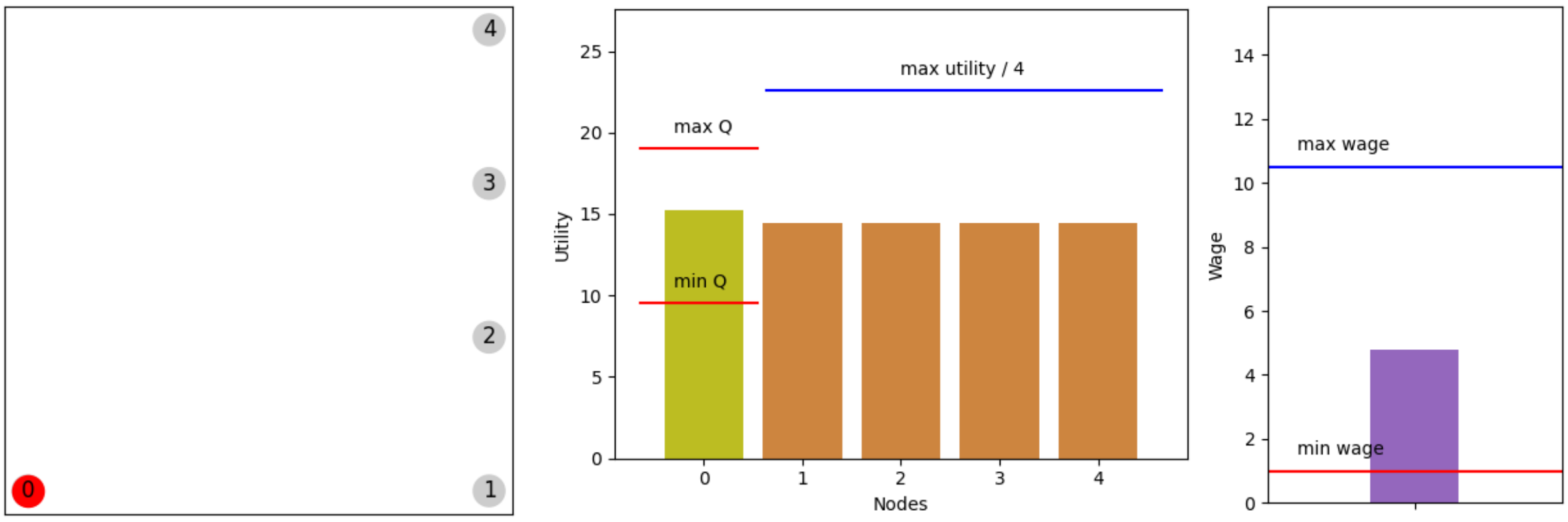

**Figure 11.** *Landowner Game, free system. Node zero (red) is the landowner: his/her utility (the olive green bar) can oscillate between max Q (perfect competition) and min Q (monopoly).*

In a free system of four peasants and a landowner (see *Figure 11*) we obviously have the same sub-optimal result as for the classic Cournot model. Workers compete with each other: each one offers more work for less money. But the limited number of workers prevents perfect competition, so neither the landowner reaches his/her maximum utility.
The wage is well below the “max wage” line.

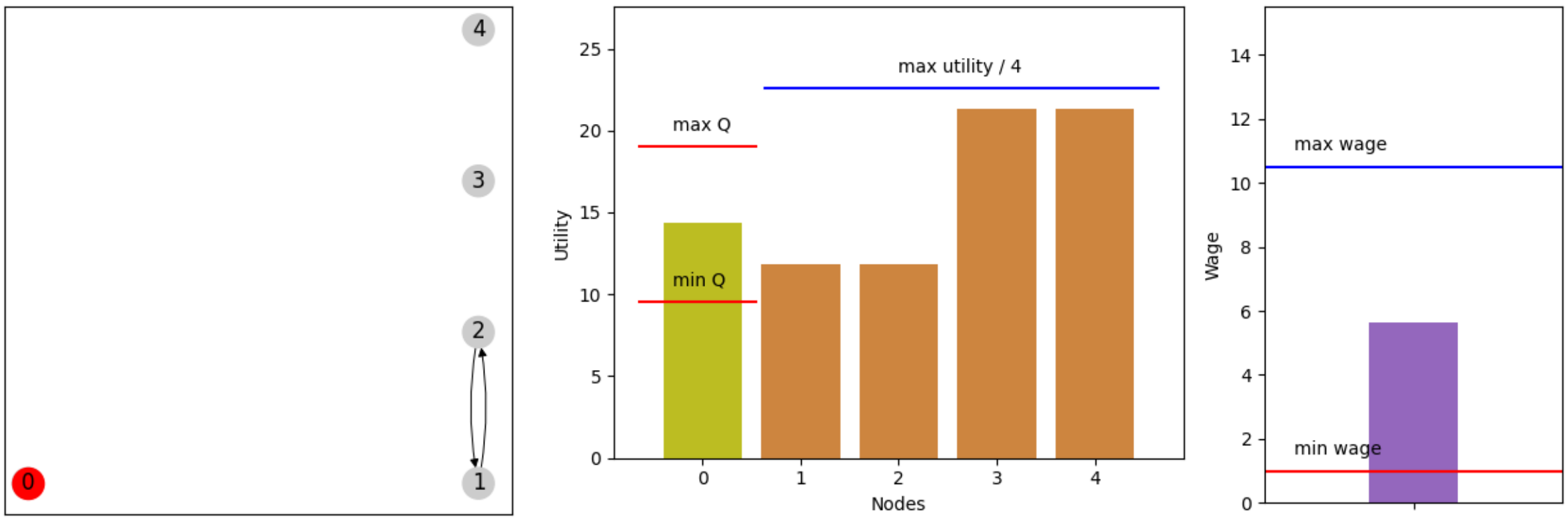

**Figure 12.** *Landowner Game, mutualism of only two nodes. If the union does not control all the farmers, then those who join lose. The wage is higher than in the free system and the quantity of work is smaller (therefore also the utility of the landowner is lower) (edges weight 0.8).*

Now suppose that a union is started (*Figure 12*), but that this organization includes only two individuals out of four. Quite surprisingly, those who join lose.

Imagine the workers striking to obtain a wage increase. If only two go on strike, wages increase because the quantity of work decreases, but all the profits go to the workers who do not strike, therefore they act as free riders.

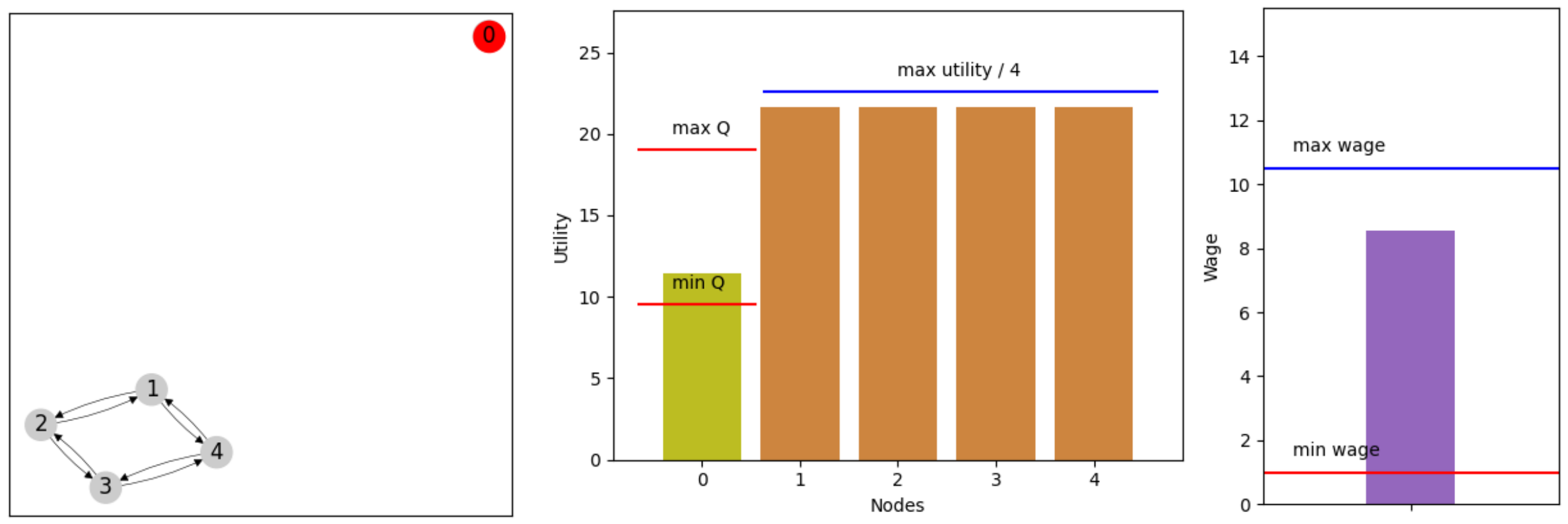

**Figure 13.** *Landowner Game, mutualism of all farmers. The union now has a monopoly on labor supply; the salary is maximum (edges weight 0.4).*

If, on the other hand, all the nodes join the union, as in *Figure 13*, the wage would reach its maximum and the amount of work would be minimal. In the middle graph, the olive bar indicating the amount of labor and the landlord's utility is near its minimum ("min Q" in *Figure 13b*), because the union now monopolizes the labor supply.

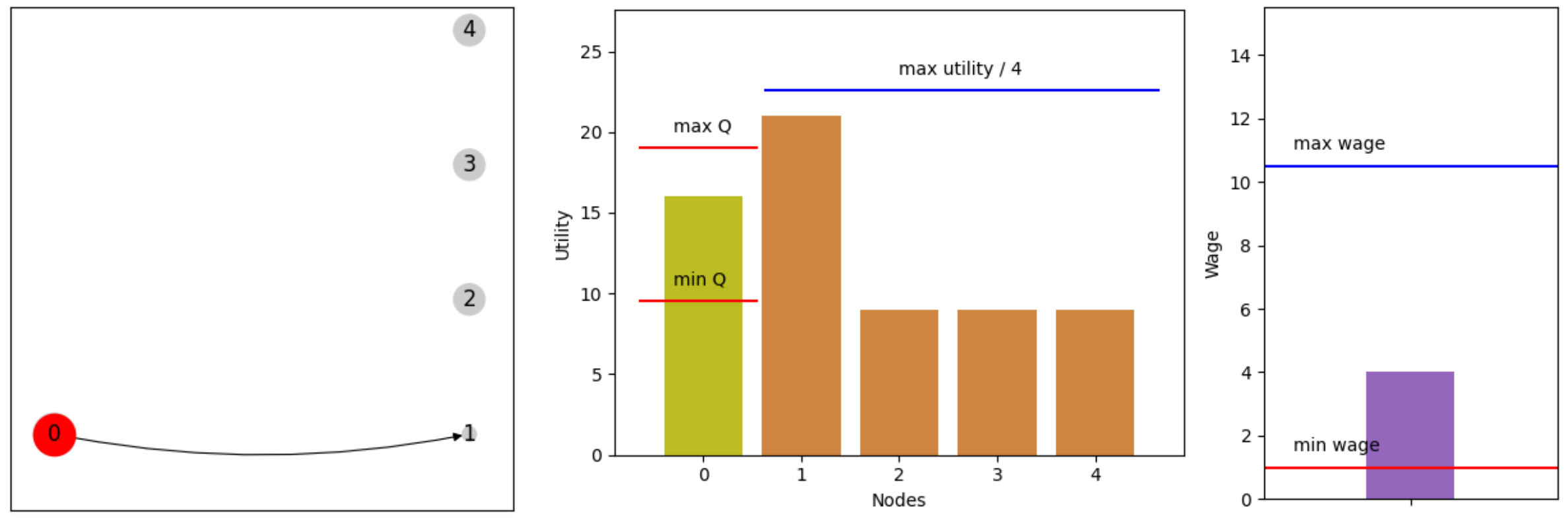

**Figure 14.** *Landowner Game, the landowner controls a farmer. The wage is lower than in the free system. The submissive peasant earns more than the others (edge weight 0.8).*

Suppose now that a peasant submits to the landowner. In doing so, the peasant would work more, making everyone's wages go down. But all in all, despite the drop in wages, the submissive peasant wins, as shown in *Figure 14*.

We are faced with a counterintuitive case, where submitting to someone implies an increase in utility for the submitted. Here obedience seems opportune.

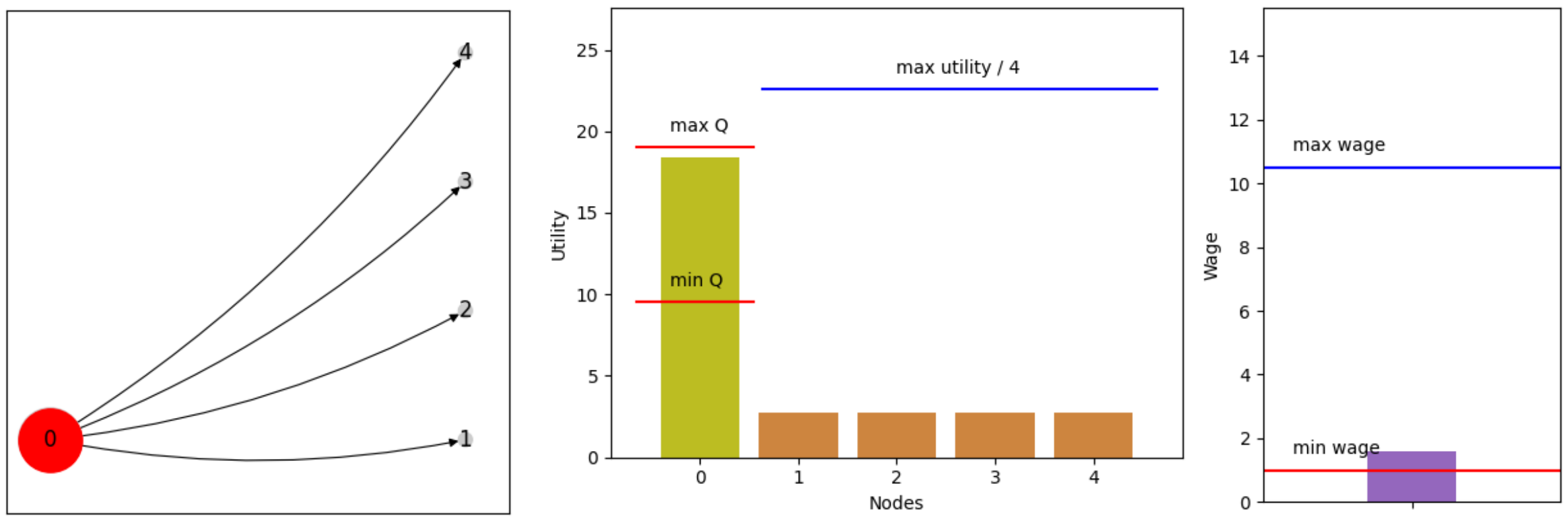

**Figure 15.** *Landowner Game, hierarchy dominated by the landowner. The wage is close to that of the perfect competition (edges weight 0.8).*

Nevertheless behind this apparent advantage there lies a trap for the peasants. In fact, if everyone acts like node (1) and submits to the landowner, the gain of (1) evaporates. All workers will be poorer, as in *Figure 15*. It is a Prisoner's Dilemma situation: individually, it is better for each farmer to submit, but the submission of everyone worsens the conditions for all. In this case the wage is close to the minimum that farmers need to cover the cost of an hour of work. On the other hand, the amount of work (represented by the green column) is the maximum.

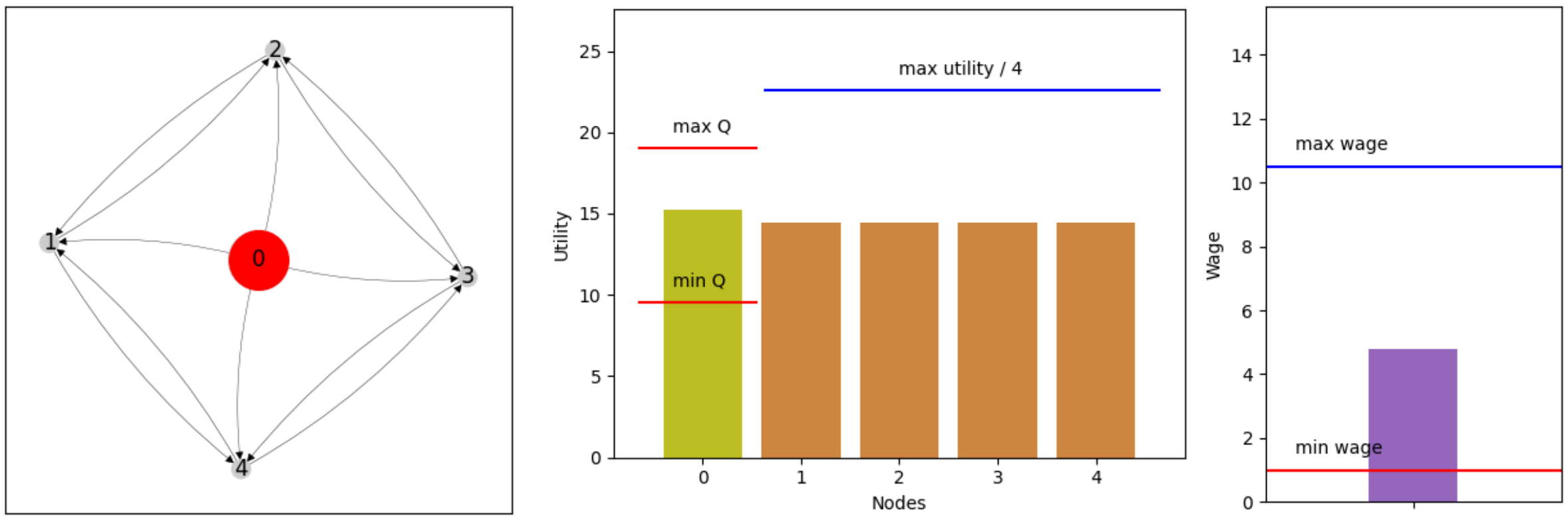

**Figure 16.** *Landowner Game, two powers that collide: a union of the peasants and a dominion of the landowner. The resulting equilibrium is the same as in the free system.*

Finally, the landowner can counterbalance the power of the union (*Figure 16*). The power of the union and that of the landowner cancel each other out as two equal and opposite forces. From the point of view of power structures, faced with a union of peasants, the landowner has two options: break the union link or increase control over the workers.

The Landowner Game shows an important property of power systems: the effect of power depends on the relationships among the inertial utilities of the nodes. When two dominant nodes have inertial utilities that cancel each other, a servant of two masters can act as a free player. That could also happen when the dominant node is in turn dominated by another node with opposite utility. In some cases, the existence of a chain can consequently improve the servant's condition.

*14. Decisions and purposes*

Whatever way it is made, a decision is always based on an individual: it is only *someone*'s decision. But the purpose of the decision and the particular utility function associated with it can take any form. There is no reason why we should restrict the set of utility functions by prohibiting, for example, two nodes from sharing the same purpose[16]. A rational agent should, by definition, be free to have any purpose, even a self-defeating[17] or contradictory one, such as the purpose of having no purpose[18]. Furthermore, decisions are objective, even though they are individual and can be referred to as *facts* of the world. Goals, on the other hand, are subjective, even though they can be general, and they concern the sphere of *language about facts*.

A human being raised outside a human community would not be a rational agent in the strict sense[19] - not having the possibility of learning the language with which to justify his/her decisions. So the very existence of rational agents attests to the possibility of dependent goals.

Furthermore, how can one ascertain the property of a purpose? For the sake of argument, one could assume that no goal ever belongs to the subject of a decision. How about an infinite chain of goals? In fact, a power structure has the same result in terms of node payoff even if each inertial goal, instead of belonging to the node, is projected onto a higher passive node, thus distancing and alienating the goals from those who pursue them. A passive node can therefore represent a rational agent, but also an intermediate or ultimate goal, an ideal or a religious belief[20].

---

[16] It has often been claimed that even altruism can be explained as an indirect self-interest, typically in the form of reputation. “But these models fail to explain two facts about human cooperation: that it takes place in groups far larger than the immediate family, and that both in real life and in laboratory experiments, it occurs in interactions that are unlikely to be repeated, and where it is impossible to obtain reputational gains from cooperating.” Bowles, Samuel and Gintis, Herbert (2011).

[17] “Preferences could be altruistic or even masochistic.” Bowles, Samuel and Gintis, Herbert (2011): 9.

[18] "There can then be nothing irrational about consistently pursuing any end whatever. As Hume extravagantly observed, he might be criticized on many grounds if he were to prefer the destruction of the entire universe to scratching his finger, but his preference could not properly be called irrational, because (contra Kant) rationality is about means rather than ends." Binmore, K. (2009): 5. See also Edvardsson, Karin and Hansson, Sven Ove (2005), where the authors try to extend rationality to pursuitability of goals.

[19] Yet, it is remarkable true that, historically, a “valuable qualitative preliminary description of the behavior of the individual is offered by the Austrian School, particularly in analyzing the economy of the isolated “Robinson Crusoe.”” Von Neumann, John and Morgenstern, Oskar (2007): 2.1.2. The very invention of Game Theory can be seen as a giant step in the attempt to overcome Robinson Crusoe’s model.

[20] In general, for a model of the type described here to make sense it is sufficient that only one of the nodes is a player, that is, capable of making choices; all others could simply serve to explain the origin of its compound utility function. In other words, this model can also be seen as a tool for generating utility functions or for analyzing them internally, dividing them into simpler components (inertial functions).

*15. Final considerations*

Thus far we have seen some of the changes brought about by the systems of power to the equilibria of games. They depend on the relationships among different inertial utilities in a game and they can have unexpected or counterintuitive outcomes.

The two main assumptions of the model are the following: 1) the transitivity of influence and 2) the definition of power as action for the sake of another individual.

1) The idea that the relationship of influence/colonization is transitive, although quite natural, can legitimately give rise to doubts. It cannot be denied that transitivity is a concrete possibility. However, it is easy to find examples of non-transitive power relations. But this should not be sufficient to exclude transitivity from the basic properties of a power model. It is clear that real power systems are extremely complex and stratified. Furthermore, there is no need to think that a concrete individual participates in just one of these systems, as if all power relations were on the same level. The absence of transitivity, where necessary, can therefore be represented in the model through multigraphs, which separate the connections into different and more or less independent levels. In this way the greater or lesser transitivity of a multisystem can take the form of a system of systems.

2) The definition of power from which we started can be described as *behaviorist*, because it is based on the visible effects of power on individuals. This is obviously a reductive view, which nonetheless has notable success in contemporary social sciences. In his influential essay *The Concept of Power,* Robert Dahl gave this famous definition[21]: “A has power over B to the extent that he can get B to do something that B would not otherwise do." Michael Taylor[22] seems to follow Dahl's assertion when he says that power is "the ability to affect the incentives facing others so that it becomes rational for them to pursue a certain course of action.” According to Pranab Bardhan[23], in terms of game theory, "an inclusive way of defining power may be to say that A has power over B if A has the ability to alter the game (preferences, strategies, sets or sets of information) in such a way that B's equilibrium outcome changes." Randall Bartlett[24] says, in turn, that power is “the ability of one actor to alter the decisions made and/or welfare experienced by another actor relative to the choices that would have been made and/or welfare that would have been experienced had the first actor not existed or acted.”

All these definitions may agree with the idea that the concept of power can be expressed through the compound utility functions, as they alter welfare, preferences and strategies.

[21] Dahl, Robert A. (1957): 202-3.
[22] Taylor, Micheal (1982): 13.
[23] Bardhan, Pranab (1991): 274.
[24] Bartlett, Randall (2006): 30.

However, what all the definitions above lack is to specify the *direction* of the game changes. It may not be bold to think that the direction is the welfare of the dominant. A definition consistent with the system outlined above will be for instance: A has power over B if A can get B to do something for the sake of A, something that B would not otherwise do.

Samuel Bowles'[25] asymmetric definition takes into account this aspect too: "For agent A to have power over agent B it is sufficient that, by imposing or threatening to impose sanctions on B, A is capable of affecting B's actions in ways which further A's interests, while B lacks this capacity with respect to A."

Surely, the concept and praxis of power include many aspects that the behaviorist definitions deliberately forget. Goldman, Alvin I. (1972), for example, identifies two main problems related to the behaviorist approach: the difficulty of measuring the actions of the individuals involved and the non-overlap between the behaviorist definition and the normal notion of power in some extreme situations.

Although there are innumerable other definitions of social power, some of them cannot help but presume that in the end the concrete effect of the power relationship consists in a change in the decisions of at least one of the individuals involved. This change must be useful for another individual and, without postulating such an effect, the very concept of power would be lost in a cloud. The hope is that the other aspects of power will be better illuminated once these assumptions have been carefully explored.

[25] Bowles, Samuel (1998): 11.

*Bibliography*

Allingham, Michael G. (1975), "Economic power and values of games," *Zeitschrift für Nationalökonomie/Journal of Economics* H. 3/4: 293-299.

Bardhan, Pranab (1991), "On the concept of power in economics," *Economics & Politics* 3.3: 265-277.

Bartlett, Randall (2006), *Economics and Power*, Cambridge Books.

Binmore, Ken (2009), *Interpersonal comparison of utility*, University College London.

Binmore, Ken (2010), "Social norms or social preferences?," *Mind & Society* 9: 139-157.

Bowles, Samuel, et al. (1998), *The politics and economics of power*, Routledge.

Bowles, Samuel and Gintis, Herbert (2011), *A Cooperative Species*, Princeton University Press.

Boyer, Marcel and Moreaux, Michel (1986), "Perfect competition as the limit of a hierarchical market game," *Economics Letters* 22.2-3: 115-118.

Czégel, Dániel and Palla, Gergely (2015), "Random walk hierarchy measure: What is more hierarchical, a chain, a tree or a star?," *Scientific reports* 5.1: 1-14.

Dahl, Robert A. (1957), "The concept of power," *Behavioral Science*, 2.3: 201-215.

Diefenbach, Thomas (2013), *Hierarchy and organisation: Toward a general theory of hierarchical social systems*, Routledge.

Dowding, Keith, ed. (2011), *Encyclopedia of power*, Sage Publications.

Edvardsson, Karin and Hansson, Sven Ove (2005), "When is a goal rational?" *Social Choice and Welfare* 24.2: 343-361.

Fehr, Ernst and Charness, Gary (2023), "Social preferences: fundamental characteristics and economic consequences".

Flache, Andreas, et al. (2017), "Models of social influence: Towards the next frontiers." *Journal of Artificial Societies and Social Simulation* 20.4.

French Jr, John RP (1956), "A formal theory of social power," *Psychological review* 63.3: 181.

Goldman, Alvin I. (1972), "Toward a theory of social power," *Philosophical Studies: An International Journal for Philosophy in the Analytic Tradition* 23.4: 221-268.

Holt, Charles A. (2019), *Markets, games, and strategic behavior: An introduction to experimental economics*, Princeton University Press.

List, John A. (2009), "Social preferences: Some thoughts from the field." *Annu. Rev. Econ.* 1.1: 563-579.

Mones, Enys, et al. (2012), "Hierarchy measure for complex networks," *PloS one* 7.3: e33799.

Moutsinas, Giannis, et al. (2021), "Graph hierarchy: a novel framework to analyse hierarchical structures in complex networks," *Scientific Reports* 11.1: 13943.

Peterson, Martin, ed. (2015), *The Prisoner's Dilemma*, Cambridge University Press.

Shapley, L. S., and Shubik, Martin (1954), "A Method for Evaluating the Distribution of Power in a Committee System," *The American Political Science Review* 48.3: 787-792.

Taylor, Michael (1982), *Community, Anarchy and Liberty*, Cambridge University Press.

Von Neumann, John and Morgenstern, Oskar (2007), *Theory of games and economic behavior*, Princeton university press.

Wiese, Harald (2009), "Applying cooperative game theory to power relations," *Quality & quantity* 43: 519-533.

# APPENDIX

| | |
|---|---|
| ***Definitions*** | System: $\Gamma = (V, \pi)$<br>Nodes set: $V = \{v_1, v_2, \dots, v_n\}$<br>Power relation: $\pi: V \times V \to R, \quad (v_i, v_j) \mapsto f_{i,j} \quad (i \neq j)$<br>Colonization matrix: $C \in R^{n,n}, C = (c_{i,j})$ |
| ***Axiom 1*** | $f_{i,j} \geq 0$ |
| ***Axiom 2*** | $\sum_{k=1}^{n} f_{k,i} < 1$ |
| ***Axiom 3*** | $\sum_{k=1}^{n} c_{k,i} = 1$ |
| ***Axiom 4*** | $c_{i,j} = \sum_{k=1}^{n} f_{k,j} c_{i,k} \quad (i, k \neq j)$ |
| ***Lemma 0*** | Colonizations are always zero or positive; $\forall i, j : c_{i,j} \geq 0$ |
| *Proof* | For *axiom 1*: $f_{i,j} \geq 0$. All the terms at the right in *axiom 4* are therefore zero or positive. |
| ***Theorem 1*** | All nodes are (partially) free; $\forall i : c_{i,i} > 0$.<br>(Reflexivity of colonization) |
| *Proof* | $$\sum_{k=1,k\neq i}^{n} c_{k,i} = \sum_{k=1,k\neq i}^{n} \sum_{j=1}^{n} f_{j,i} c_{k,j} = \sum_{j=1}^{n} \sum_{k=1,k\neq i}^{n} f_{j,i} c_{k,j}$$ for *axiom 3*: $\sum_{k=1}^{n} c_{k,i} = 1$, then, for *lemma 0*, $\sum_{k=1,k\neq i}^{n} c_{k,j} \leq 1$ and<br>$\sum_{j=1}^{n} \sum_{k=1,k\neq i}^{n} f_{j,i} c_{k,j} = \sum_{j=1}^{n} f_{j,i} \sum_{k=1,k\neq i}^{n} c_{k,j} \leq \sum_{j=1}^{n} f_{j,i}$.<br>For *axiom 2*: $\sum_{k=1}^{n} f_{k,i} < 1$, then $$\sum_{k=1,k\neq i}^{n} c_{k,i} \leq \sum_{k=1}^{n} f_{k,i} < 1$$ and<br>$c_{i,i} > 0$. |
| ***Lemma 1*** | A node is totally free if, and only if, it has no predecessors;<br>$\forall x : (c_{x,x} = 1 \Leftrightarrow \forall y \neq x : f_{y,x} = 0)$. |
| *Proof* | (⇐) For *axiom 4*, $\forall y \neq x : c_{y,x} = 0$. So for *axiom 3*, $c_{x,x} = 1$.<br>(⇒) From $c_{x,x} = 1$ follows for *axiom 3* and *lemma 0* that<br>$\forall y \neq x : c_{y,x} = 0$.<br>For *axiom 4*, $\forall y \neq x, c_{y,x} = \sum_{k=1}^{n} f_{k,x} c_{y,k} = 0$.<br>All the elements of the sum must be zero, for they cannot be negative: |
| | $\forall k\,,\,\forall y\,\neq\,x : f_{k,x}c_{y,k}\,=\,0.$<br>For *theorem 1*: $c_{i,i}\,>\,0$ for every $i$. Therefore, every element with $k\,=\,y$ will be positive in every colonization of $i$ unless:<br>$\forall y\,,y\,\neq\,x : f_{y,x}\,=\,0.$ |
| ***Theorem 2*** | A node $i$ colonizes a node $j$ if, and only if, there exists at least one path[26] from $i$ to $j$;<br>$\forall i, \forall j\,\neq\,i : (\,c_{i,j}\,>\,0 \Leftrightarrow \exists P(i,j)\,).$<br>(Transitivity of colonization) |
| *Proof* | (⇒) $c_{i,j}\,>\,0$ means that $j$ is colonized by $i$. So, for *axiom 3*, $j$ is not totally free. For *lemma 1*, $j$ must at least have an edge from a node $k$ with $c_{i,k}\,>\,0$, and so on. The recursive formula of *axiom 4* only stops when it reaches the node $i$, with $c_{i,i}\,>\,0$ (*theorem 1)*.<br>(⇐) Vice versa: if there exists a path from $i$ to $j$ that passes by nodes $a,\,\cdots,\,y,\,z$, for *axiom 4:*<br>$c_{i,j}\,\geq\,f_{z,j}c_{i,z}$<br>$c_{i,z}\,\geq\,f_{y,z}c_{i,y}$<br>…<br>$c_{i,a}\,\geq\,f_{i,a}c_{i,i}.$<br>All the edges $f$ must have a positive weight, for they belong to the path. It follows then for *theorem 1* that: $c_{i,a}\,>\,0,\,\cdots,c_{i,z}\,>\,0\,,c_{i,j}\,>\,0.$ |
| ***Definition 1 (binary mutualism)*** | $mutualism(i,j)\,=\,2\cdot min(c_{i,j}, c_{j,i})$ |
| ***Definition 2 (binary cooperation)*** | $cooperation(i,j)\,=\,c_{i,j}\,+\,c_{j,i}$ |
| ***Definition 3 (binary*** | $hierarchy(i,j)\,=\,\left\|c_{i,j}\,-\,c_{j,i}\right\|$ |

[26] A path is an ordered set P(a,z) = ( f(a,b), f(b,c), f(c,d), ... , f(y,z) ) of consecutive positive-weighted edges from a to z.

| | |
|---|---|
| ***hierarchy)*** | |
| ***Theorem 3*** | $hierarchy(i,j) = cooperation(i,j) - mutualism(i,j)$ |
| *Proof* | $mutualism(i,j) = 2 \cdot min(c_{i,j}, c_{j,i}) = c_{i,j} + c_{j,i} - \lvert c_{i,j} - c_{j,i} \rvert$ |
| ***Definition 4 (mutualism)*** | $mutualism(\Gamma) = \frac{2 \sum_{\{i,j\} \in V_2} min(c_{i,j}, c_{j,i})}{n-1}$ |
| ***Definition 5 (cooperation)*** | $cooperation(\Gamma) = \frac{\sum_{\{i,j\} \in V_2} (c_{i,j} + c_{j,i})}{n-1}$ |
| ***Definition 6 (hierarchy)*** | $hierarchy(\Gamma) = \frac{\sum_{\{i,j\} \in V_2} \lvert c_{i,j} - c_{j,i} \rvert}{n-1}$ |
| ***Theorem 4*** | $hierarchy(\Gamma) = cooperation(\Gamma) - mutualism(\Gamma)$ |
| *Proof* | $cooperation(\Gamma) - mutualism(\Gamma)$ $= \frac{\sum_{\{i,j\} \in V_2} (c_{i,j} + c_{j,i})}{n-1} - \frac{2 \sum_{\{i,j\} \in V_2} min(c_{i,j}, c_{j,i})}{n-1}$ $= \frac{\sum_{\{i,j\} \in V_2} (c_{i,j} + c_{j,i} - 2min(c_{i,j}, c_{j,i}))}{n-1}$ $= \frac{\sum_{\{i,j\} \in V_2} (c_{i,j} + c_{j,i} - (c_{i,j} + c_{j,i} - \lvert c_{i,j} - c_{j,i} \rvert))}{n-1}$ $= \frac{\sum_{\{i,j\} \in V_2} \lvert c_{i,j} - c_{j,i} \rvert}{n-1} = hierarchy(\Gamma).$ |
| ***Definition 7 (freedom)*** | $freedom(\Gamma) = \frac{\sum_{i=1}^{n} c_{i,i} - 1}{n-1} = \frac{n - \sum_{\{i,j\} \in V_2} (c_{i,j} + c_{j,i}) - 1}{n-1}$ <br> $= 1 - \frac{\sum_{\{i,j\} \in V_2} (c_{i,j} + c_{j,i})}{n-1}$ <br> $= 1 - cooperation(\Gamma)$ |
| ***Theorem 5*** | In a Prisoner's Dilemma with payoffs: Cooperates-Cooperates (p, p), Cooperates-Defects (q, r), Defects-Cooperates (r, q), Defects-Defects (s, s), if the average of r and q is between s and p, the mutualism sufficient to shift the equilibrium to Cooperates-Cooperates is $2 \cdot max(\frac{r-p}{r-q}, \frac{s-q}{r-q})$ . |
| *Proof* | The payoff's relation $r > p > s > q$ put the game in the Defects-Defects equilibrium. To shift the equilibrium one must have: <br> $p > c_{1,1}q + c_{2,1}r > s$ <br> $p > c_{2,2}q + c_{1,2}r > s$ <br><br> And, if for instance, $c_{1,2} < c_{2,1}$[27]: <br> $c_{1,2} = \frac{1}{2}mutualism(\Gamma)$ <br> $c_{2,1} = \frac{1}{2}mutualism(\Gamma) + hierarchy(\Gamma)$ <br> $c_{2,2} = 1 - c_{1,2}$ <br> $c_{1,1} = 1 - c_{2,1}$. <br><br> Combining the previous formulas one obtains that if <br> $p < \frac{r+q}{2} < s$ <br> and <br> $mutualism(\Gamma) > 2 \cdot max(\frac{r-p}{r-q}, \frac{s-q}{r-q})$ |

[27] **Γ** is a graph with two nodes. Let us suppose that c(1,2) < c(2,1), otherwise one can change the variables.

| | |
|---|---|
| | the equilibrium shifts. |
| ***Theorem 6*** | In a Prisoner's Dilemma there always exists a level of hierarchy that can shift the equilibrium to Defects-Cooperates or Cooperates-Defects. |
| *Proof* | Hierarchy can move the perceived payoff of a player as close as you want towards the other. It means that the payoff at r and q for a player can always swap to the point where equilibrium shifts. |